\documentclass[runningheads]{llncs}

\usepackage[T1]{fontenc}

\usepackage{graphicx}
\graphicspath{ {figures/} }

\usepackage{amsmath}
\usepackage{booktabs}
\usepackage{xurl}
\usepackage{hyperref}
\hypersetup{hidelinks}

\usepackage{nicematrix}
\usepackage{tikz}
\usetikzlibrary{arrows.meta, positioning, fit, backgrounds, shapes.geometric}

\usepackage{pgfplots}
\pgfplotsset{compat=1.18}

\usepackage{enumitem}
\setlist[itemize]{noitemsep, topsep=0pt}

\begin{document}

\title{Small, Free, and Effective: Orchestrating Open-Weight Small Language Models to Outperform Single LLM for Malware Analysis}
\titlerunning{Orchestrating SLMs to Outperform Single LLM for Malware Analysis}
\authorrunning{A.~ElZemity et al.}

\author{Adel ElZemity\orcidID{0000-0002-5402-7837} 
\and Shujun Li\orcidID{0000-0001-5628-7328} 
\and Budi Arief\orcidID{0000-0002-1830-1587}}

\institute{University of Kent, Canterbury, United Kingdom\\
\email{\{ae455, s.j.li, b.arief\}@kent.ac.uk}}
\maketitle

\begin{abstract}
Malware analysis demands rapid interpretation of complex detonation reports spanning filesystem, network, and process behaviours. While large language models (LLMs) demonstrate impressive capabilities for technical artifact interpretation, the opacity and escalating API costs of closed-weight frontier models motivate exploration of open-weight alternatives. However, many open-weight models are themselves large, demanding significant compute resources and incurring non-trivial hosting costs that place them beyond reach for resource-constrained deployments. This paper investigates whether orchestrated ensembles of small language models (SLMs) can match or exceed single LLM performance on structured questions about malware detonation reports. We established baselines by testing eleven open-weight SLMs, three cyber security pre-trained models, and six frontier LLMs on Meta's CyberSecEval Malware Analysis benchmark. We then designed and evaluated four orchestration architectures: (i) a multi-agent pipeline that decomposes analysis into structured evidence-collection and reasoning stages, (ii) an adversarial debate framework in which two agents iteratively critique each other's reasoning, (iii) a hierarchical consultation system that pairs a general-purpose SLM with a cyber-specialised expert model, and (iv) a hybrid architecture that combines evidence-grounded pipelines with adversarial debate reasoning. The hybrid system (Qwen3-4B with Foundation-Sec-8B) achieved 35.30\% overall accuracy, exceeding the strongest cyber-specialised baseline (22.54\%) and the strongest ungrounded frontier baseline (34.77\%); when given the same evidence pipeline, grounded Gemini remained the strongest configuration at 38.22\%. Case studies on malware from the wild (UNC5142 and Lumma Stealer) illustrated the hybrid system's ability to correct reasoning errors on novel evasion techniques such as EtherHiding and ClickFix. These findings show that evidence-grounded orchestration can substantially improve the performance of collaborative SLMs for supporting interpretation of malware detonation reports.
\end{abstract}

\section{Introduction}

Malware analysis remains a critical bottleneck in cyber security operations. Analysts must rapidly triage suspicious samples, interpret complex detonation reports spanning filesystem modifications, network communications, and process behaviours, and assess threat severity under time pressure~\cite{Akhtar2023Evaluation,Aslan2021Intelligent,Mat2024A}. Traditional static and dynamic analysis pipelines generate rich telemetry---dynamic analysis through sandboxed execution environments and static analysis through disassembly, string extraction, and structural inspection---but extracting actionable intelligence from these multi-faceted reports demands expert knowledge of malware techniques, operating system internals, and attack frameworks such as MITRE ATT\&CK~\cite{strom2018mitre}. As malware sophistication and volume continue to escalate, the need for automated assistance that can comprehend, reason about, and summarise behavioural evidence has become acute~\cite{Akhtar2023Evaluation,Catak2020Deep}.

Recent advances in large language models (LLMs) have demonstrated LLMs' capabilities for interpreting technical artifacts and answering domain-specific questions~\cite{deason2025}. However, leading closed-weight frontier models remain opaque and costly to deploy via API~\cite{Hou2023Large}, raising concerns about data privacy, vendor lock-in~\cite{Raza2025Industrial}, and the reproducibility of security-critical decisions~\cite{Li2025Security}. At the same time, the open-weight model ecosystem has matured rapidly~\cite{leon2025gpt}, spanning models from compact, consumer-deployable sizes to large parameter counts that rival closed-weight systems in resource demands. Within this spectrum, increasingly capable small language models (SLMs) have emerged as viable alternatives for specialised tasks without the compute and cost overhead of their larger open-weight counterparts. Following Belcak et al.~\cite{belcak2025smalllanguagemodelsfuture}, we use SLM operationally to describe models deployable locally under our fixed single-GPU budget. Our evaluated SLMs contain at most 8B parameters and run at 4-bit quantisation on one RTX~4090; this is a study-specific deployment criterion rather than a definitive SLM--LLM boundary.

While individual SLMs often lag behind frontier LLMs on complex reasoning tasks, orchestration strategies (such as multi-agent systems, debate frameworks, and hierarchical consultation patterns) offer a path to improve SLMs' collective capabilities. Multi-agent architectures have shown promise in decomposing complex tasks into specialised subtasks~\cite{wu2023autogen}, debate-style interactions can expose reasoning flaws and improve answer quality~\cite{du2025improving}, and hierarchical consultation allows general-purpose models to seek targeted expertise~\cite{lin2024soen101codegenerationemulating}. Open-weight SLMs such as \textit{Mistral}, \textit{Phi}, \textit{Qwen}, and \textit{Llama} variants can be self-hosted, audited, and fine-tuned for domain specificity, while cyber security pre-trained models (ranging from SLM to LLM scale) offer specialised knowledge of malware techniques and defensive concepts~\cite{2025llm}. Furthermore, ensembles provide a path to fuse diverse inductive biases while offering defence-in-depth through redundancy~\cite{Sagi2018Ensemble,Mienye2022A}. Nonetheless, the extent to which orchestrated SLM ensembles can close the performance gap to frontier LLMs on malware analysis tasks remains an open question and it deserves a detailed exploration.

Malware analysts increasingly demand transparent reasoning, reproducible artifacts, and explicit risk controls~\cite{Saqib2024A,Manthena2024Explainable,Fan2020Can} when interpreting detonation reports, identifying malicious behaviours, and assessing threat severity~\cite{Mahmoud2024Redefining,Or-Meir2019Dynamic}. These requirements are better served by open-weight models than closed-weight alternatives: weights can be inspected and audited~\cite{Saqib2024A,Manthena2024Explainable}, local deployment enables greater reproducibility through seed and environment control, and self-hosting allows organisations to implement their own risk control policies over the inference pipeline. However, these advantages only translate into operational value if orchestrated SLM systems can deliver competitive accuracy on real-world malware analysis benchmarks. As such, this paper asks: \textit{Under a fixed single-GPU budget, can orchestrated open-weight SLMs match or exceed single LLMs when answering structured questions about malware detonation reports?} Note that we evaluate report comprehension and behavioural interpretation, rather than static binary reverse engineering or autonomous malware analysis.

This work makes the following key \textbf{contributions}:
\begin{enumerate}
\item \textbf{A systematic evaluation of three orchestration architectures} (agentic, debate, and consult) that instantiate complementary hypotheses for amplifying SLMs' capabilities, evaluated on the CyberSecEval Malware Analysis benchmark across 11 open-weight SLMs and 3 cyber-specialised models. 

\item \textbf{An empirical characterisation of the interaction between grounding and debate.} Component ablations show that ungrounded debate can introduce drift on retrieval-oriented questions, while evidence grounding stabilises these exchanges and preserves peer-critique gains on harder behavioural-reasoning questions.

\item \textbf{An empirical evidence} showing that grounded orchestration benefits both SLMs and LLMs. The open-weight SLM hybrid achieved 35.30\% accuracy, compared with 34.77\% for the strongest ungrounded frontier baseline, while grounded Gemini achieved 38.22\%. This demonstrates that collaborative SLMs can substantially narrow, but do not eliminate, the evidence-matched performance gap.

\item \textbf{Qualitative case studies} on malware samples from the wild (UNC5142 EtherHiding campaign and Lumma Stealer with ClickFix), illustrating cases in which the hybrid system is capable of correcting \emph{reasoning errors} on novel evasion techniques that defeat single-model approaches.
\end{enumerate}

The rest of this paper is organised as follows. Section~\ref{sec:related} reviews related work in different areas. Section~\ref{sec:methodology} describes our methodology, including the four orchestration architectures and the evaluation benchmark. Section~\ref{sec:results} presents experimental results, comparing single-model baselines with orchestrated systems and reporting ablation studies. Section~\ref{sec:further_discussions} discusses the implications of our findings for operational deployment. Finally, Section~\ref{sec:conclusion} concludes our paper and provides several directions for future work.

\section{Related Work}
\label{sec:related}

Our work builds on four converging research threads: LLMs for cyber security and malware analysis, multi-agent LLM systems, debate frameworks for improving LLM reasoning, and the emerging capabilities of SLMs.

\subsection{LLMs for Cyber Security and Malware Analysis}

LLMs have been increasingly applied to cyber security tasks, with recent literature documenting their use across vulnerability detection~\cite{ccetin2024empirical}, malware analysis and network intrusion detection~\cite{Xu2025Large}, and threat intelligence~\cite{Yigit2025Generative}. In particular, LLMs have demonstrated promising potential for interpreting malicious artifacts: for instance, Patsakis et al.~\cite{PATSAKIS2024124912} showed that LLMs achieved 69.56\% accuracy in extracting malicious URLs from obfuscated code in real-world campaigns like Emotet, while outperforming symbolic analysis in bypassing common evasion techniques. Al-Karaki et al.~\cite{maniriho2024malwaredetection} presented a comprehensive framework for LLM-based malware detection, identifying key challenges including dataset limitations and the need for domain-specific fine-tuning. The CyberSecEval benchmark suite~\cite{bhatt2024cyberseceval3} provides standardised evaluation protocols for assessing LLM capabilities on security tasks, including the malware analysis benchmark we use in this study. While these papers have shown that LLMs can assist with security analysis, they primarily evaluated single models in isolation; our work extends this line by investigating whether orchestration can improve compact-model' performance on answering questions related to detonation-report.

\subsection{Multi-Agent LLM Systems}

Multi-agent architectures decompose complex tasks across specialised LLM agents that communicate and collaborate. Wu et al.~\cite{wu2023autogen} introduced AutoGen, a framework enabling customisable agents with flexible conversation patterns for tasks spanning coding, mathematics, and decision-making. A subsequent work showed that multi-agent orchestration can provide value through deterministic quality and consistency rather than speed alone~\cite{feng2025multiagentsurvey}. Liu et al.~\cite{liu2023dynamic} proposed dynamic agent networks that optimise team composition based on task requirements, while hierarchical frameworks such as AgentOrchestra~\cite{zhang2026} use central planning agents that delegate to specialised sub-agents. These systems have shown success in software development, question answering, and enterprise operations. Our agentic and consult systems build on these principles. Related security systems include API-call-based malware frameworks~\cite{Belaoued2020MACoMal,Qaisar2021A,Uysal2025A}, CVE-Genie for vulnerability reproduction~\cite{ullah2025cvegenie}, Chimera for insider-threat simulation and log generation~\cite{yu2026chimera}, and SentinelOne's adversarial-consensus pipeline for tool-assisted binary analysis~\cite{stokes2026consensus}. In contrast, we evaluate retrieval, consultation, and debate for question answering over existing detonation reports.

\subsection{LLM Debate Frameworks}

Debate-style orchestration, where multiple LLM agents critique each other's reasoning, has emerged as an effective technique for improving factuality and complex reasoning. Du et al.~\cite{du2025improving} demonstrated that multi-agent debate improves mathematical and strategic reasoning by exposing flaws through adversarial exchange. Recent extensions include the Mixture-of-Agents framework~\cite{wang2024mixtureofagents}, which organises proposer and aggregator agents in structured layers to achieve state-of-the-art results using open-source models, and adaptive heterogeneous debate~\cite{adaptivedebate}, which was shown to have achieved 4--6\% accuracy gains over standard debate through dynamic agent weighting. Chen et al.~\cite{chen2023reconcile} showed that round-table consensus among diverse LLMs can improve reasoning on complex benchmarks. These works established that structured debate improves reasoning quality, but also revealed a trade-off: excessive debate rounds can introduce tangential information that degrades performance on straightforward questions. Our hybrid system addresses this limitation by grounding debate agents in systematically collected evidence, preventing the drift phenomenon while preserving the benefits of peer critique.

\subsection{Small Language Models}

The capabilities of SLMs -- typically defined as models deployable on consumer hardware with acceptable latency~\cite{belcak2025smalllanguagemodelsfuture} -- have advanced rapidly. Lu et al.~\cite{lu2025smalllanguagemodelssurvey} surveyed SLMs in the 100M--5B parameter range, documenting competitive performance on common sense reasoning, mathematics, and domain-specific tasks when compared to much larger models. Recent work~\cite{Zhang2025Rise} demonstrated that well-chosen SLMs can outperform frontier LLMs including GPT-4 variants in specific use cases, particularly when enhanced through fine-tuning, prompt engineering, or ensemble techniques. Recent work on CyberPal~2.0 develops cybersecurity-expert SLMs ranging from 4B to 20B parameters for threat-intelligence and investigation tasks~\cite{levi2025cyberpal}. Our work complements model-development research by evaluating cyber-specialised compact models as orchestration partners for detonation-report question answering.

\section{Methodology}
\label{sec:methodology}

Our methodology evaluates whether orchestration improves the performance of open-weight SLMs on structured detonation-report question answering relative to single LLMs. We began by establishing baseline performance: we tested a diverse collection of models (including general-purpose open-weight SLMs, frontier LLMs, and cyber security pre-trained models) as solo agents on the CyberSecEval Malware Analysis benchmark~\cite{bhatt2024cyberseceval3}. We selected this benchmark as, to the best of our knowledge, it is the only benchmark designed to automate the evaluation of language models specifically for malware analysis. This benchmark exercises Hybrid Analysis~\cite{hybridanalysis} detonation reports\footnote{In this work, we take malware analysis detonation reports rather than malware binaries as the input, a standard practice in dynamic malware analysis~\cite{Or-Meir2019Dynamic,Mahmoud2024Redefining}.} through multi-topic, multi-difficulty multiple-choice questions, providing strict accuracy metrics stratified by difficulty tier (Easy, Medium, Hard). The solo-model baselines provided reference points for measuring orchestration gains, and they established which model architectures and parameter scales performed the best on malware analysis tasks when operating independently.

We tested eleven general-purpose open-weight SLMs spanning 0.6B to 8B parameters: Qwen3-0.6B~\cite{hf_qwen3_0_6b}, Llama-3.2-1B~\cite{hf_llama_3_2_1b}, Qwen2.5-1.5B-Instruct~\cite{hf_qwen2_5_1_5b_instruct}, DeepSeek-R1-Distill-Qwen-1.5B~\cite{hf_deepseek_r1_qwen_1_5b}, SmolLM2-1.7B~\cite{hf_smolm2_1_7b}, Phi-3.5-mini-instruct (3.8B)~\cite{hf_phi_3_5_mini_instruct}, Gemma-3-4B-IT~\cite{hf_gemma_3_4b_it}, Qwen3-4B~\cite{hf_qwen3_4b}, Qwen2.5-Coder-7B-Instruct~\cite{hf_qwen2_5_coder_7b_instruct}, Ministral-8B~\cite{hf_ministral_8b}, and Llama-3.1-8B-Instruct~\cite{hf_llama_3_1_8b_instruct}. Their performances were compared against those of three open-weight cyber security pre-trained models spanning SLM to LLM scale (DeepHat-V1-7B~\cite{hf_deephat_v1_7b}, Foundation-Sec-8B-Instruct~\cite{hf_foundation_sec_8b_instruct}, and Llama-Primus-Nemotron-70B~\cite{nvidia2025nemotron}), and six frontier LLMs (Gemini 3 Pro Preview~\cite{google2025gemini3}, Claude Opus 4.5~\cite{anthropic2025claude}, GPT-5.2~\cite{openai2025gpt5}, DeepSeek V3.2~\cite{liu2025deepseek}, Llama 4 Scout, and Llama 4 Maverick~\cite{meta2025llama4}).

Open-weight models were sourced from Hugging Face; closed-weight frontier models (Gemini 3 Pro Preview, Claude Opus 4.5, GPT-5.2) were accessed via their official APIs with default sampling parameters. To assess potential contamination, we conducted temporal, reference-corpus overlaps, and answer-probing checks for Foundation-Sec-8B (Appendix~\ref{app:contamination}). All SLM inference experiments were conducted on a single NVIDIA RTX 4090 GPU (24\,GB VRAM); the hybrid system required approximately 6\,GB VRAM with 4-bit quantisation.

Single-pass LLMs might fail on malware analysis tasks through three characteristic failure modes observed during baseline evaluation. First, \emph{context overload}: full Hybrid Analysis JSON reports commonly exceed model context windows, forcing truncation of critical evidence (network telemetry, extracted payloads). Second, \emph{surface-level pattern matching}: models frequently classify samples based on conspicuous keywords (e.g., blockchain terms $\rightarrow$ mining, Chrome overlays $\rightarrow$ phishing) rather than tracing causal execution chains. Third, \emph{domain knowledge gaps}: correct interpretation of specific MITRE ATT\&CK techniques requires specialised knowledge that general-purpose pre-training does not consistently provide. The four architectures address these failure modes complementarily: the agentic pipeline addresses context overload through structured retrieval; the debate system addresses surface-level reasoning through peer critique; the consult system addresses domain gaps through on-demand expert access; and the hybrid combines all three interventions.

After establishing solo-model baselines, we designed and implemented four distinct orchestration architectures: a specialised multi-agent pipeline (agentic system), an adversarial debate framework (debate system), a hierarchical consultation system (consult system), and a hybrid architecture that combines evidence-grounded pipelines with adversarial debate reasoning (hybrid system). Each architecture embodies a different hypothesis for capability amplification: specialisation through task decomposition, peer critique through adversarial reasoning, expert guidance through hierarchical consultation, and synergistic combination of evidence collection with structured debate. We then evaluated these orchestration systems by running representative SLMs through each architecture on the same malware analysis benchmark. Finally, we compared the orchestrated system performance against the solo-model baselines to quantify the performance gains attributable to orchestration, and to identify which architectural patterns provide the largest improvements across different difficulty tiers and model sizes. Model-internal Mixture-of-Experts~\cite{Cai2024A} routing is outside our scope, which focuses on inference-time orchestration of independently deployable models.

\begin{figure}[!t]
\centering
\resizebox{\linewidth}{!}{%
\begin{tikzpicture}[
    font=\large,
    box/.style={rectangle, rounded corners, draw, thick, align=center, minimum height=0.85cm, minimum width=2.1cm, inner sep=3.2pt},
    io/.style={trapezium, trapezium left angle=75, trapezium right angle=105, draw, thick, align=center, minimum height=0.85cm, minimum width=2.1cm, inner sep=3.2pt},
    agent/.style={box},
    tool/.style={box},
    debate/.style={box},
    phasebox/.style={rectangle, draw, rounded corners, inner sep=11pt},
    arrow/.style={-Latex, thick, shorten >=1.5pt, shorten <=1.5pt},
    dblarrow/.style={Latex-Latex, thick, shorten >=1.5pt, shorten <=1.5pt},
    node distance=0.5cm and 0.4cm
]

\node[io, minimum width=1.8cm] (loop) at (-5.5, 1.5) {Hybrid Analysis\\Report \& Question};

\node[agent, right=0.8cm of loop] (ingest) {\textbf{Ingestion Agent}\\Chunks report};

\draw[arrow] (loop.east) -- (ingest.west);

\node[box, above=1.2cm of ingest, xshift=2.4cm] (enrich) {\textbf{Enrichment}\\MITRE context};

\draw[arrow] (ingest.north) |- (enrich.west);

\node[tool, below=1.2cm of ingest, xshift=2.4cm] (toolpath) {\textbf{Tool Search}\\grep/jq};

\draw[arrow] (ingest.south) |- (toolpath.west);

\node[agent, right=2.0cm of ingest] (miner) {\textbf{Evidence Miner}\\Select \& extract};

\draw[arrow] (ingest.east) -- (miner.west);
\draw[arrow] (enrich.east) -| (miner.north);
\draw[arrow] (toolpath.east) -| (miner.south);

\node[io, right=1.0cm of miner, minimum width=1.8cm] (evidence) {\textbf{Evidence}\\Bundle};

\draw[arrow] (miner.east) -- (evidence.west);

\node[debate, right=1.0cm of evidence, yshift=0.8cm, minimum width=2.5cm] (agentA) {\textbf{Agent A}\\(General SLM)};
\node[debate, right=1.0cm of evidence, yshift=-0.8cm, minimum width=2.5cm] (agentB) {\textbf{Agent B}\\(Cyber Expert)};

\begin{scope}[on background layer]
\node[phasebox, fit=(agentA)(agentB), inner sep=10pt] (debatebox) {};
\end{scope}

\draw[arrow] (evidence.east) -- ++(0.25,0) |- (agentA.west);
\draw[arrow] (evidence.east) -- ++(0.25,0) |- (agentB.west);

\draw[dblarrow] (agentA.south) -- node[right, font=\small, xshift=1pt] {N rounds} (agentB.north);

\node[agent, below=0.9cm of debatebox.south] (verifier) {\textbf{Verifier Agent}\\Cross-check};

\draw[arrow] (debatebox.south) -- (verifier.north);

\node[io, left=1.0cm of verifier] (output) {\textbf{Final}\\Answer};

\draw[arrow] (verifier.west) -- (output.east);

\begin{scope}[on background layer]
\node[rectangle, rounded corners, draw, dashed, thick, fit=(loop)(ingest)(enrich)(toolpath)(miner), inner sep=12pt] {};
\node[rectangle, rounded corners, draw, dashed, thick, fit=(evidence)(debatebox), inner sep=12pt] {};
\end{scope}

\end{tikzpicture}%
}
\caption{\textbf{Hybrid orchestration architecture}. \emph{Left (evidence collection phase):} The four-stage agentic pipeline shared with the standalone agentic system (Section~\ref{sec:agentic})---ingestion, enrichment, tool-search, and evidence mining---extracts and validates supporting evidence. The standalone agentic system terminates after a single reasoning and verifier step using one model. \emph{Right (debate reasoning phase):} Unique to the hybrid---Agent~A (general-purpose SLM) and Agent~B (cyber-specialised model) engage in $N$ rounds of evidence-grounded structured debate, followed by a verifier that validates the final answer against the evidence bundle.}
\label{fig:orchestration}
\end{figure}
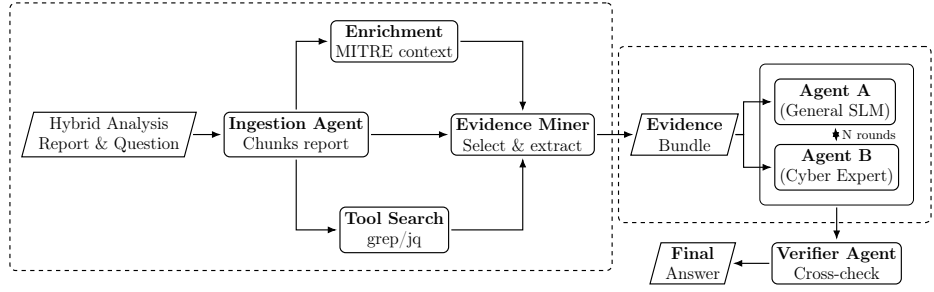

\subsection{Agentic System}
\label{sec:agentic}

The first architecture uses a task-oriented multi-agent workflow for interpreting detonation reports. The agentic pipeline is shown in the left panel of Figure~\ref{fig:orchestration}. The system decomposes the analysis process into six specialised stages orchestrated by a central controller. An \textit{ingestion agent} prepares the workspace by parsing and structuring the relevant malware report: the Hybrid Analysis JSON dossier is divided into semantically coherent sections (process inventory, network telemetry, filesystem modifications, registry changes, and extracted strings) rather than fixed-length windows, preserving field-level context. Each section is chunked at a maximum of 512 tokens with 64-token overlap between adjacent chunks to prevent boundary truncation of multi-field artefacts. (Here ``retrieval'' refers to selecting relevant chunks from the already-provided report JSON, not fetching from external sources.) An \textit{enrichment agent} augments this data by fetching relevant MITRE ATT\&CK technique descriptions~\cite{kuppa2021linking}. The enrichment trigger policy is heuristic-driven: the agent scans ingested chunks for ATT\&CK technique identifiers (e.g., \texttt{T1059}, \texttt{T1204}) and for a curated vocabulary of 150 technique-indicative keywords (e.g., ``persistence'', ``lateral movement'', ``credential dumping''), a size determined empirically to balance enrichment coverage against false trigger rate. When a match is detected, the agent queries the MITRE ATT\&CK STIX API~\cite{strom2018mitre} for the corresponding technique and appends it to the shared evidence state; if no match is found, the enrichment phase is skipped. The enrichment decision is deterministic (keyword/pattern match), not LLM/SLM-driven, to ensure reproducibility. To locate specific indicators of compromise, a \textit{tool-search agent} generates and executes sandboxed read-only search commands (\texttt{grep} for string/pattern matching and \texttt{jq} for structured JSON field extraction) against the on-disk report file. The agent receives the benchmark question and the list of ingested chunk summaries as input, generates candidate search queries, executes them in a restricted shell with no network access, and appends raw output to the shared state. An \textit{evidence miner} then extracts supporting snippets from the accumulated report content and assigns each snippet a confidence score $\tau_i \in [0,1]$, computed as the cosine similarity between the snippet's embedding and the question embedding using \texttt{all-MiniLM-L6-v2}~\cite{allMiniLML6v2}, a lightweight (22.7M parameter) sentence embedding model designed for semantic similarity and information retrieval, chosen for its computational efficiency and widespread validation in the community. Snippets with $\tau_i > \tau = 0.65$ are retained in the evidence bundle, a threshold determined empirically to balance evidence recall against noise. These retrieved artefacts are synthesised by a \textit{reasoning agent} to formulate an answer, which is finally subjected to a quality gate by a \textit{verifier agent} before being returned. All agents operate on a shared state object that ensures every decision can be traced back to its supporting evidence.

\subsection{Debate System}
\label{sec:debate}

The second architecture pairs two SLM agents (Agent~A and Agent~B) in a structured adversarial debate. For each question, the agents engage in $N$ rounds: in each round, each agent receives the original prompt plus the full debate history, critiques the opponent's reasoning, and produces a revised rationale and a structured answer. After each round, a controller checks whether the agents agree; if they disagree, it may retrieve and provide relevant MITRE ATT\&CK technique descriptions to inform the next round. Once all rounds are complete, the controller produces a final answer by reviewing the full debate transcript. Motivated by security systems that use dedicated verification or sceptical peer review~\cite{stokes2026consensus,ullah2025cvegenie}, the debate tests whether explicit criticism exposes reasoning weaknesses missed in single-pass evaluation.

\subsection{Consult System}

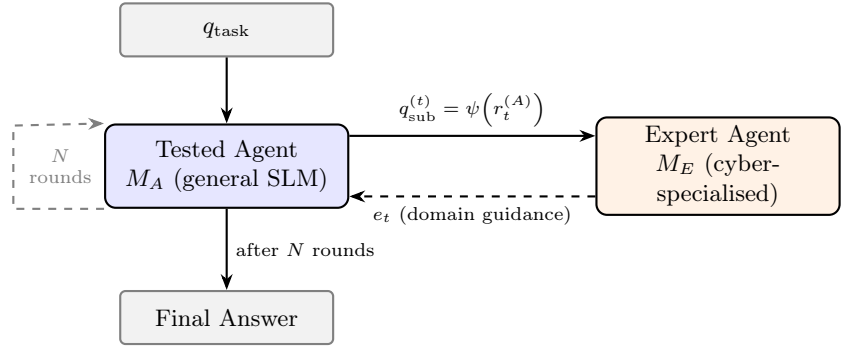
\begin{figure}[t]
\centering
\begin{tikzpicture}[
    box/.style   = {rectangle, rounded corners=4pt, draw=black, thick,
                    fill=blue!10, text width=3.0cm, align=center,
                    minimum height=1.1cm, font=\small},
    expbox/.style= {rectangle, rounded corners=4pt, draw=black, thick,
                    fill=orange!10, text width=3.0cm, align=center,
                    minimum height=1.1cm, font=\small},
    iobox/.style = {rectangle, rounded corners=2pt, draw=gray, thick,
                    fill=gray!10, text width=2.6cm, align=center,
                    minimum height=0.7cm, font=\small},
    arr/.style   = {-{Stealth[length=6pt]}, thick},
    darr/.style  = {-{Stealth[length=6pt]}, thick, dashed},
    lbl/.style   = {font=\scriptsize, midway}
]

\node[iobox]  (input)  at (0,   5.0) {$q_{\text{task}}$};
\node[box]    (tested) at (0,   3.2) {Tested Agent\\$M_A$ (general SLM)};
\node[expbox] (expert) at (6.5, 3.2) {Expert Agent\\$M_E$ (cyber-specialised)};
\node[iobox]  (output) at (0,   1.2) {Final Answer};

\draw[arr] (input.south) -- (tested.north);

\draw[arr] ([yshift=4mm]tested.east) -- ([yshift=4mm]expert.west)
    node[lbl, above] {$q_{\text{sub}}^{(t)} = \psi\!\left(r_t^{(A)}\right)$};

\draw[darr] ([yshift=-4mm]expert.west) -- ([yshift=-4mm]tested.east)
    node[lbl, below] {$e_t$ (domain guidance)};

\draw[darr, gray]
    (tested.south west) -- ++(-1.2, 0) -- ++(0, 1.1) -- (tested.north west);
\node[font=\scriptsize, text=gray, align=center] at (-2.2, 3.2) {$N$\\rounds};

\draw[arr] (tested.south) -- (output.north)
    node[lbl, right] {after $N$ rounds};

\end{tikzpicture}
\caption{\textbf{Consult system architecture.} The tested agent ($M_A$, general-purpose SLM) receives the full benchmark prompt $q_{\text{task}}$ and iteratively consults the stateless expert agent ($M_E$, cyber-specialised), which never sees $q_{\text{task}}$. Each round, $\psi$ extracts a sub-question $q_{\text{sub}}^{(t)}$ from the tested agent's rationale; the expert returns focused domain guidance $e_t$, which accumulates in $H_E$ across $N$ rounds. The tested agent then produces the final answer.}
\label{fig:consult}
\end{figure}

The third architecture pairs a ``tested'' general-purpose SLM with a pre-trained cyber security expert model in a hierarchical consultation loop, illustrated in Figure~\ref{fig:consult}. The tested agent owns the task (receives the full benchmark prompt) and may pose at most one sub-question per round; the expert sees only the extracted question and responds with focused domain guidance. An extraction function $\psi$ isolates the sub-question in two stages: it first looks for explicit markers (``Question:'', ``Query:''), then falls back to the first interrogative sentence longer than 10 words, a minimum length heuristic to exclude trivially short or incomplete questions. Expert guidance accumulates across rounds, allowing the tested agent to iteratively refine its answer without the expert ever controlling the task. This design probes whether on-demand domain expertise can close the performance gap of compact SLMs lacking deep malware knowledge.

We formalise the consult interaction as a hierarchical loop. Let $q_{\text{task}}$ be the full benchmark prompt, $M_A$ the tested agent model, and $M_E$ the expert model. At round $t$, the tested agent $A$ produces a rationale $r_t^{(A)}$ and potentially a specific query $q_{\text{sub}}$ for the expert. An extraction function $\psi$ isolates the explicit question from the tested agent's output. The expert agent $E$ (which does not see $q_{\text{task}}$) provides a domain-specific explanation $e_t$. The tested agent then updates its state using the accumulated history of expert advice $H_E = \left\{ \left(q_{\text{sub}}^{(i)}, e_i\right) \right\}_{i=1}^t$:
\begin{equation}
q_{\text{sub}}^{(t)} = \psi\left(r_t^{(A)}\right);\;
e_t = M_E\left(q_{\text{sub}}^{(t)}\right);\;
r_{t+1}^{(A)} = M_A\left(q_{\text{task}}, H_E\right)
\end{equation}
This formalisation highlights $M_E$ as a ``stateless oracle'' relative to the main task, distinguishing this architecture from the state-sharing debate agents.

\subsection{Hybrid System}

The fourth architecture combines the evidence retrieval capabilities of the agentic system with the adversarial reasoning of the debate system, as illustrated in Figure~\ref{fig:orchestration}, which shows the full hybrid architecture with the agentic pipeline on the left and the debate reasoning phase on the right. The hybrid system operates in three distinct phases designed to address the complementary weaknesses of its component architectures: evidence collection, adversarial debate reasoning, and final verification.

\paragraph{Phase 1: Evidence Collection.} The first four stages of the agentic pipeline execute unchanged: ingestion, enrichment, tool-search, and evidence mining (see Section~\ref{sec:agentic} for the full explanation). This phase produces the structured evidence bundle $\mathcal{B}$ formalised below, which serves as a fixed evidence context for the debate reasoning phase.

Formally, let $\mathcal{R}$ denote the raw Hybrid Analysis report (JSON dossier). The \emph{evidence bundle} $\mathcal{B}$ is the union of three extraction outputs: $\phi_{\text{chunk}}(\mathcal{R})$ (ingestion agent's text chunks), $\phi_{\text{enrich}}(\mathcal{R})$ (enrichment agent's external context, e.g., MITRE ATT\&CK descriptions), and $\text{Exec}(\phi_{\text{tool}}(\mathcal{R}))$ (results of sandboxed commands generated by the tool-search agent). The evidence miner applies a filtering function $F_\tau$ based on a confidence threshold $\tau = 0.65$ (determined empirically; see Section~\ref{sec:agentic}) to produce the final validated evidence set $E_{\text{final}}$, which serves as the fixed evidence context for the debate reasoning phase:
\begin{equation}
\mathcal{B} = \phi_{\text{chunk}}(\mathcal{R}) \cup \phi_{\text{enrich}}(\mathcal{R}) \cup \text{Exec}(\phi_{\text{tool}}(\mathcal{R}))
\end{equation}
\begin{equation}
E_{\text{final}} = \{ e \in \mathcal{B} \mid \text{Confidence}(e) > \tau \}
\end{equation}

\paragraph{Phase 2: Debate Reasoning.} Rather than passing evidence to a single reasoning agent, the hybrid system instantiates two debate agents that receive the collected evidence alongside the original question. For instance, Agent~A is a general-purpose SLM (Qwen3-4B), selected based on its strong baseline performance across all orchestration systems; Agent~B is a cyber-specialised model (Foundation-Sec-8B), selected to provide complementary domain expertise while maintaining capacity balance (within 2$\times$ parameter ratio). The agents engage in $N$ rounds of structured debate following the protocol described in Section~\ref{sec:debate}, but with a critical modification: agents are explicitly instructed to cite collected evidence when defending their positions and to challenge claims that lack evidential support.

We model the debate as a Markov process over $t$ rounds. In round $t$, Agent~A's response $r_t^{(A)}$ is conditional on the original question $q$, the evidence bundle $E_{\text{final}}$, the debate history $H_{t-1}$, and the opponent's previous argument $r_{t-1}^{(B)}$:
\begin{equation}
r_t^{(A)} = M_A \left( q, E_{\text{final}}, H_{t-1}, r_{t-1}^{(B)} \right)
\end{equation}
Unlike standard debates, the hybrid system enforces a \emph{grounding constraint} via the verifier: a response $r$ is valid if and only if every claim $c \in r$ maps to a supporting snippet in $E_{\text{final}}$ with cosine similarity $S(c, e)$ (computed using the same embedding model as $\tau$) exceeding threshold $\lambda = 0.55$, set lower than $\tau$ since claims in a debate response are often paraphrases or inferences from the evidence rather than direct matches, requiring a more relaxed similarity criterion:
\begin{equation}
\text{Valid}(r) \iff \forall c \in r,\; \exists e \in E_{\text{final}} \text{ s.t.\ } S(c, e) \geq \lambda
\end{equation}
This constraint prevents the ``drift'' phenomenon observed in pure debates, where agents introduce tangential information that degrades easy-questions' accuracy.

\paragraph{Phase 3: Verification.} After the debate concludes, the verifier agent validates the debate conclusion against the evidence bundle, checking that the selected answer has supporting evidence with confidence above $\lambda$ and that the reasoning chain is internally consistent. If verification fails, the system falls back to the answer most directly supported by the evidence bundle, bypassing the debate conclusion to ensure that the final output is always grounded in collected evidence.

\section{Results}
\label{sec:results}

This section presents the experimental findings of our study. Throughout, the task context is that of a Security Operations Centre (SOC) analyst querying a malware detonation report --- the primary real-world setting in which such reports are interpreted under time pressure: given a Hybrid Analysis JSON dossier, can the system correctly answer structured questions about the sample's persistence mechanisms, network behaviour, and evasion techniques? We begin with single-model baselines across various parameter scales to establish single-model reference points. We then detail the performance of the four orchestrated architectures (agentic, debate, consult, and hybrid) to quantify the capability gains achieved through different ensemble strategies. We conclude with qualitative case studies of real-world malware samples from the wild, demonstrating the hybrid system's ability to identify and reason through novel evasion techniques.

\subsection{Benchmark and Evaluation Protocol}

We ground our experiments in the CyberSecEval Malware Analysis benchmark (CyberSOCEval test suite)~\cite{deason2025}, which pairs Hybrid Analysis detonation reports with multi-topic, multi-difficulty multiple-choice questions spanning evidence retrieval, behavioural interpretation, risk scoring, and system-interaction audits. The multi-label format (up to nine answer options per question) penalises both omissions and incorrect option selections, providing exact-match accuracy and Jaccard partial-credit scores stratified by topic, difficulty tier, and malware family. We reserve \emph{hallucination} for rationales containing claims unsupported by the supplied report or enrichment evidence. Ground-truth options are isolated from model inputs, so systems must locate and reconcile the relevant report fragments rather than memorise expected outputs.

\paragraph{A Benchmark Example.} A typical Medium-difficulty question asks which persistence mechanisms a sample uses (multi-select from 9 options); the correct answer is embedded within thousands of lines of process telemetry. A single-pass LLM must locate and cross-reference the relevant report fragments without retrieval; the agentic system uses \texttt{jq} and \texttt{grep} to retrieve them directly.

Let $D = \{\text{Easy}, \text{Medium}, \text{Hard}\}$ denote the difficulty tiers, $N_d$ the number of questions in Tier $d$, and $Acc_d$ the model's accuracy on that tier; $N_{\text{total}} = 609$ ($N_{\text{Easy}}=451$, $N_{\text{Med}}=136$, $N_{\text{Hard}}=22$). The \emph{overall weighted accuracy} is calculated as follows:
\begin{equation}
\text{Acc}_{\text{overall}} = \sum_{d \in D} \frac{N_d}{N_{\text{total}}} \cdot \text{Acc}_d = \frac{451 \cdot \text{Acc}_{\text{Easy}} + 136 \cdot \text{Acc}_{\text{Med}} + 22 \cdot \text{Acc}_{\text{Hard}}}{609}
\end{equation}
Because Easy questions constitute 74\% of the dataset, performance on this tier dominates the overall score, i.e., a model's ability to handle straightforward evidence-retrieval queries has a greater impact on its aggregate ranking than proficiency on the rare, complex Hard questions. This weighting must be kept in mind when interpreting overall accuracy figures: a system that excels on Hard questions but degrades on Easy ones may rank lower overall, even if it demonstrates superior reasoning capability on the most challenging cases. We therefore report per-tier accuracy alongside overall scores throughout this section.

\paragraph{Model Selection for Orchestration Experiments.} Qwen3-4B is selected as the general-purpose agent based on its leading performance across all solo and orchestrated configurations; Foundation-Sec-8B provides cyber-specialised expertise within the optimal capacity balance. The three selection criteria (capacity balance, complementary expertise, strong baseline) are discussed in Section~\ref{sec:further_discussions}.

\begin{table}[t]
\centering
\caption{Accuracy percentages for each model on the Malware Analysis benchmark, stratified by difficulty level.}
\label{tab:mal_ana_stats}
\small
\resizebox{\textwidth}{!}{\begin{tabular}{lccccc}
\toprule
Model & Params & Easy ($n=451$) & Medium ($n=136$) & Hard ($n=22$) & Overall ($n=609$)\\
\midrule
\multicolumn{6}{@{}l}{\textbf{LLMs}} \\
Llama 4 Scout          & 109B & 25.50\% & 14.00\% & 0.00\% & 22.01\%\\
Llama 4 Maverick      & 400B & 31.00\% & 20.50\% & 13.64\% & 28.03\%\\
DeepSeek V3.2         & 685B & 33.50\% & 22.00\% & 13.64\% & 30.21\%\\
Claude Opus 4.5        & --- & 36.25\% & 24.75\% & 21.59\% & 33.15\%\\
Gemini 3 Pro Preview   & --- & \textbf{38.00\%} & \textbf{26.00\%} & \textbf{22.73\%} & \textbf{34.77\%}\\
GPT-5.2                & --- & 34.50\% & 23.50\% & 20.45\% & 31.54\%\\
\midrule
\multicolumn{6}{@{}l}{\textbf{Cyber security language models}}\\
DeepHat-V1-7B         & 7B & 16.41\% & 11.03\% & 4.55\% & 14.78\%\\
Foundation-Sec-8B     & 8B & \textbf{22.65\%} & \textbf{13.45\%} & \textbf{4.55\%} & \textbf{19.96\%}\\
Llama-Primus-Nemotron-70B & 70B & 24.78\% & 18.00\% & 4.55\% & 22.54\%\\
\midrule
\multicolumn{6}{@{}l}{\textbf{SLMs}}\\
Qwen3-0.6B            & 0.6B & 12.94\% & 6.61\% & 0.00\% & 11.05\%\\
Llama-3.2-1B          & 1B & 10.17\% & 5.88\% & 0.00\% & 8.87\%\\
Qwen2.5-1.5B-Instruct & 1.5B & 11.72\% & 7.35\% & 0.00\% & 10.34\%\\
DeepSeek-R1-Distill-Qwen-1.5B & 1.5B & 12.42\% & 8.09\% & 4.55\% & 11.17\%\\
SmolLM2-1.7B          & 1.7B & 10.86\% & 6.62\% & 0.00\% & 9.52\%\\
Phi-3.5-mini-instruct & 3.5B & 15.74\% & 10.29\% & 4.55\% & 14.12\%\\
Gemma-3-4B-IT         & 4B & 16.62\% & 11.03\% & 4.55\% & 14.94\%\\
Qwen3-4B              & 4B & \textbf{18.40\%} & \textbf{12.50\%} & 4.55\% & \textbf{16.58\%}\\
Qwen2.5-Coder-7B-Instruct & 7B & 12.70\% & 9.19\% & 4.55\% & 11.66\%\\
Ministral-8B          & 8B & 11.75\% & 8.82\% & 4.55\% & 10.84\%\\
Llama-3.1-8B-Instruct & 8B & 13.24\% & 9.56\% & 4.55\% & 12.15\%\\
\bottomrule
\end{tabular}}
\end{table}

\begin{table}[t]
\centering
\caption{Comparison of orchestrated systems against single-model baselines on the Malware Analysis benchmark. Orchestrated systems are tested with four representative SLMs: Qwen3-0.6B, Phi-3.5-mini-instruct (3.5B), Qwen3-4B, and Ministral-8B. \emph{Agentic} system uses a single SLM with sandboxed command-line tools; \emph{Debate} pairs each SLM with Foundation-Sec-8B (seven rounds); \emph{Consult} uses each SLM as tested agent consulting Foundation-Sec-8B as cyber expert (seven rounds); \emph{Hybrid} combines agentic evidence collection with debate reasoning (seven rounds).\label{tab:systems_comparison}}
\small
\resizebox{\textwidth}{!}{\begin{tabular}{llcccc}
\toprule
System & Model & Easy ($n=451$) & Medium ($n=136$) & Hard ($n=22$) & Overall ($n=609$)\\
\midrule
\multicolumn{6}{l}{\textbf{Single-model baselines}}\\
& Best open-weight SLM (Qwen3-4B) & 18.40\% & 12.50\% & 4.55\% & 16.58\%\\
& Best single LLM (Gemini 3 Pro Preview) & \textbf{38.00\%} & \textbf{26.00\%} & \textbf{22.73\%} & \textbf{34.77\%}\\
\midrule
\multicolumn{6}{l}{\textbf{Agentic (with tools)}}\\
& Qwen3-0.6B & 22.62\% & 17.65\% & 9.09\% & 21.02\%\\
& Phi-3.5-mini-instruct & 24.61\% & 19.85\% & 13.64\% & 23.15\%\\
& Qwen3-4B & \textbf{26.50\%} & \textbf{21.30\%} & \textbf{20.00\%} & \textbf{25.11\%}\\
& Ministral-8B & 25.71\% & 20.59\% & 18.18\% & 24.30\%\\
\midrule
\multicolumn{6}{l}{\textbf{Debate (7 rounds, paired with Foundation-Sec-8B)}}\\
& Qwen3-0.6B & 20.00\% & 14.80\% & 13.64\% & 18.60\%\\
& Phi-3.5-mini-instruct & 22.40\% & 17.70\% & 18.18\% & 21.20\%\\
& Qwen3-4B & \textbf{25.50\%} & \textbf{19.80\%} & \textbf{22.73\%} & \textbf{24.13\%}\\
& Ministral-8B & 25.00\% & 19.10\% & \textbf{22.73\%} & 23.60\%\\
\midrule
\multicolumn{6}{l}{\textbf{Consult (7 rounds, paired with Foundation-Sec-8B)}}\\
& Qwen3-0.6B & 21.51\% & 16.18\% & \textbf{9.09\%} & 19.87\%\\
& Phi-3.5-mini-instruct & 23.06\% & 18.38\% & \textbf{9.09\%} & 21.51\%\\
& Qwen3-4B & \textbf{24.50\%} & \textbf{19.10\%} & \textbf{9.09\%} & \textbf{22.74\%}\\
& Ministral-8B & 23.95\% & 18.38\% & \textbf{9.09\%} & 22.17\%\\
\midrule
\multicolumn{6}{l}{\textbf{Hybrid (evidence-informed debate, 7 rounds)}}\\
& Qwen3-0.6B + Foundation-Sec-8B & 28.82\% & 19.85\% & 18.18\% & 26.44\%\\
& Phi-3.5-mini-instruct + Foundation-Sec-8B & 34.59\% & 24.26\% & 22.73\% & 31.86\%\\
& Qwen3-4B + Foundation-Sec-8B & \textbf{38.14\%} & \textbf{27.21\%} & \textbf{27.27\%} & \textbf{35.30\%}\\
& Ministral-8B + Foundation-Sec-8B & 36.59\% & 25.74\% & \textbf{27.27\%} & 33.83\%\\
\bottomrule
\end{tabular}}
\end{table}

\subsection{Single-Model Baselines}

Table~\ref{tab:mal_ana_stats} contrasts representative LLM and SLM baselines across benchmark difficulty tiers, establishing the empirical gap our orchestration aims to close.

The baseline profiling revealed that \emph{parameter count alone does not predict malware-analysis capability}. For instance, Qwen3-4B achieved 16.58\% overall accuracy, outperforming all tested 7--8B models, while sub-2B models cluster tightly in the 8.87--11.17\% range with minimal performance differences. 

Based on this profiling, we selected four representative SLMs for detailed orchestration experiments: Qwen3-0.6B, Phi-3.5-mini-instruct (3.5B), Qwen3-4B, and Ministral-8B. These models span four distinct size classes (sub-1B, mid-range 3.5B, mid-range 4B, and 8B), representing diverse architectural families (Qwen, Phi, and Mistral variants), and demonstrating strong baseline performance within their respective categories: Qwen3-0.6B achieved the highest overall accuracy (11.05\%) among sub-1B models, Phi-3.5-mini-instruct delivered competitive mid-tier performance (14.12\%), Qwen3-4B achieved the best overall performance among all open-weight SLMs (16.58\%), and Ministral-8B provided an 8B reference point (10.84\%). This selection enabled us to assess whether the benefits of orchestration generalise across model scales and whether architectural diversity influences ensemble effectiveness.

\subsection{Orchestrated Systems Performance}

Table~\ref{tab:systems_comparison} presents the performance of all three orchestrated systems against their single-model baselines, demonstrating the gains achieved through orchestration.

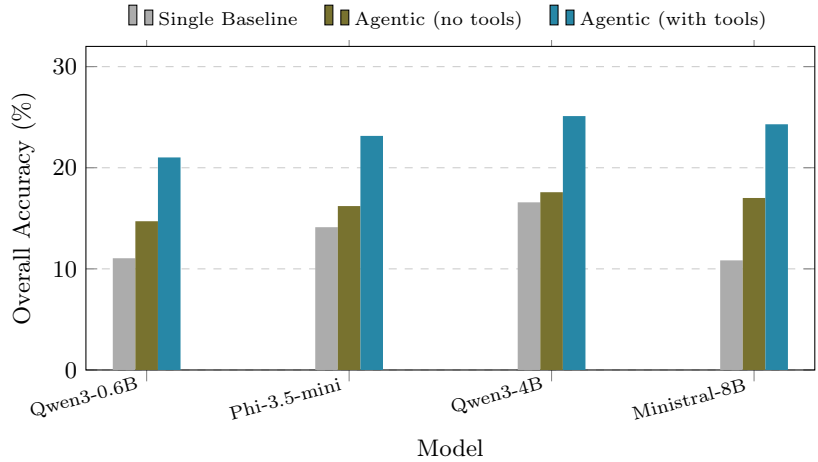
\begin{figure}[!t]
\centering
\begin{tikzpicture}
\begin{axis}[
    ybar,
    width=0.92\linewidth,
    height=0.48\linewidth,
    ymin=0, ymax=32,
    bar width=8.5pt,
    ylabel={Overall Accuracy (\%)},
    xlabel={Model},
    symbolic x coords={Qwen3-0.6B,Phi-3.5-mini,Qwen3-4B,Ministral-8B},
    xtick=data,
    x tick label style={font=\scriptsize, rotate=15, anchor=east},
    ymajorgrids=true,
    grid style={dashed, gray!40},
    legend style={
        font=\scriptsize,
        at={(0.5,1.02)},
        anchor=south,
        legend columns=3,
        draw=none,
        /tikz/every even column/.append style={column sep=8pt}
    }
]
\addplot+[draw=none, fill=gray!65, bar shift=-8.5pt, forget plot] coordinates {
    (Qwen3-0.6B,11.05) (Phi-3.5-mini,14.12) (Qwen3-4B,16.58) (Ministral-8B,10.84)
};
\addplot+[draw=none, fill=olive!70!black, bar shift=0pt, forget plot] coordinates {
    (Qwen3-0.6B,14.71) (Phi-3.5-mini,16.21) (Qwen3-4B,17.58) (Ministral-8B,17.01)
};
\addplot+[draw=none, fill=cyan!60!black, bar shift=8.5pt, forget plot] coordinates {
    (Qwen3-0.6B,21.02) (Phi-3.5-mini,23.15) (Qwen3-4B,25.11) (Ministral-8B,24.30)
};
\addlegendimage{ybar,ybar legend,draw=none,fill=gray!65}
\addlegendimage{ybar,ybar legend,draw=none,fill=olive!70!black}
\addlegendimage{ybar,ybar legend,draw=none,fill=cyan!60!black}
\legend{Single Baseline,Agentic (no tools),Agentic (with tools)}
\end{axis}
\end{tikzpicture}
\caption{The ablation analysis of the agentic system showing the impact of command-line tools across four representative SLMs. Tool access provides the largest performance boost for all models. Qwen3-4B with tools achieves the best overall performance among agentic configurations (25.11\%), surpassing all open-weight SLM solo baselines.}
\label{fig:agentic_results}
\end{figure}

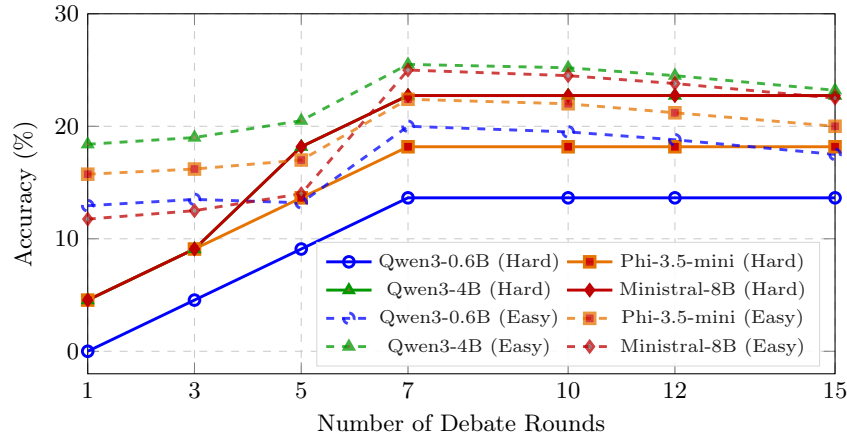
\begin{figure}[!t]
\centering
\begin{tikzpicture}
\begin{axis}[
    width=0.94\linewidth,
    height=0.52\linewidth,
    xmin=1, xmax=15,
    ymin=-2, ymax=30,
    xtick={1,3,5,7,10,12,15},
    xlabel={Number of Debate Rounds},
    ylabel={Accuracy (\%)},
    ymajorgrids=true,
    xmajorgrids=true,
    grid style={dashed, gray!40},
    legend style={font=\scriptsize, at={(0.98,0.02)}, anchor=south east, legend columns=2, fill opacity=0.9, draw=gray!30}
]
\addplot+[color=blue, mark=o, line width=1.1pt] coordinates {(1,0.00) (3,4.55) (5,9.09) (7,13.64) (10,13.64) (12,13.64) (15,13.64)};
\addplot+[color=orange!90!black, mark=square*, line width=1.1pt] coordinates {(1,4.55) (3,9.09) (5,13.64) (7,18.18) (10,18.18) (12,18.18) (15,18.18)};
\addplot+[color=green!60!black, mark=triangle*, line width=1.1pt] coordinates {(1,4.55) (3,9.09) (5,18.18) (7,22.73) (10,22.73) (12,22.73) (15,22.73)};
\addplot+[color=red!75!black, mark=diamond*, line width=1.1pt] coordinates {(1,4.55) (3,9.09) (5,18.18) (7,22.73) (10,22.73) (12,22.73) (15,22.73)};
\addplot+[color=blue, mark=o, dashed, opacity=0.75, line width=1.1pt] coordinates {(1,12.94) (3,13.50) (5,13.20) (7,20.00) (10,19.50) (12,18.80) (15,17.50)};
\addplot+[color=orange!90!black, mark=square*, dashed, opacity=0.75, line width=1.1pt] coordinates {(1,15.74) (3,16.20) (5,17.00) (7,22.40) (10,22.00) (12,21.20) (15,20.00)};
\addplot+[color=green!60!black, mark=triangle*, dashed, opacity=0.75, line width=1.1pt] coordinates {(1,18.40) (3,19.00) (5,20.50) (7,25.50) (10,25.20) (12,24.50) (15,23.20)};
\addplot+[color=red!75!black, mark=diamond*, dashed, opacity=0.75, line width=1.1pt] coordinates {(1,11.75) (3,12.50) (5,14.00) (7,25.00) (10,24.50) (12,23.80) (15,22.50)};
\legend{
Qwen3-0.6B (Hard),Phi-3.5-mini (Hard),Qwen3-4B (Hard),Ministral-8B (Hard),
Qwen3-0.6B (Easy),Phi-3.5-mini (Easy),Qwen3-4B (Easy),Ministral-8B (Easy)
}
\end{axis}
\end{tikzpicture}
\caption{The debate system's performance as a function of debate rounds (1--15 rounds), where each SLM debates with Foundation-Sec-8B. Results show consistent improvement on hard questions (solid lines) that plateaus after 7--10 rounds, while easy questions (dashed lines) exhibit clear degradation with increased rounds, declining from peak performance at rounds 7--10 to lower accuracy by round 15. Qwen3-4B achieves the highest performance across debate rounds.}
\label{fig:debate_results}
\end{figure}

\subsubsection{Agentic System Results}

The agentic multi-stage pipeline shows that tool-augmented workflows can materially upgrade SLM capabilities. Across all four representative models (Qwen3-0.6B, Phi-3.5-mini-instruct, Qwen3-4B, Ministral-8B), the agentic system consistently outperformed their single-model baselines, with ablation studies (Figure~\ref{fig:agentic_results}) indicating that access to carefully sandboxed command-line tools provided the largest incremental boost to overall performance for each model. Notably, Qwen3-4B with tools achieved 25.11\% overall accuracy, surpassing all open-weight SLM solo baselines and exceeding the weakest LLM baseline (Llama 4 Scout at 22.01\%), though it fell below the stronger frontier models, demonstrating that tool-augmented evidence retrieval alone is insufficient to close the gap to the strongest LLM baselines and that additional reasoning mechanisms --- as provided by the hybrid system --- are necessary to exceed the best single LLM baseline (Gemini 3 Pro Preview at 34.77\%).

\subsubsection{Debate System Results}

When we applied the debate-style orchestration, pairing each of the four representative SLMs with the cyber-specialised Foundation-Sec-8B model, we observed a complementary set of effects that held consistently across all tested configurations (Figure~\ref{fig:debate_results}). Increasing the number of debate rounds consistently improved performance on the \emph{hard} questions, but degraded accuracy on the easiest items; inspection of the logs showed that introducing irrelevant or tangential evidence into the debate sometimes pulled initially correct answers towards incorrect alternatives. Across all tested pairings, performances on hard questions saturated after roughly 7--10 rounds, while accuracy on easy questions continued to decline with additional rounds beyond this point. The debate system achieved overall scores that exceeded all open-weight SLM solo baselines when operated at the optimal round count, though they remain below the strongest frontier LLM baselines. Among the representative models, Qwen3-4B paired with Foundation-Sec-8B achieved the highest debate performance (24.13\% overall), demonstrating that mid-sized models with strong baseline capabilities can engage effectively in multi-round critique.

The full all-pairs debate-partner matrix and extended interpretation are provided in Appendix~\ref{app:debate_matrix} (Table~\ref{tab:debate_matrix}); in summary, results remain symmetric by role assignment, favour complementary general+cyber pairings, and suggest an association between large parameter imbalance and lower performance.

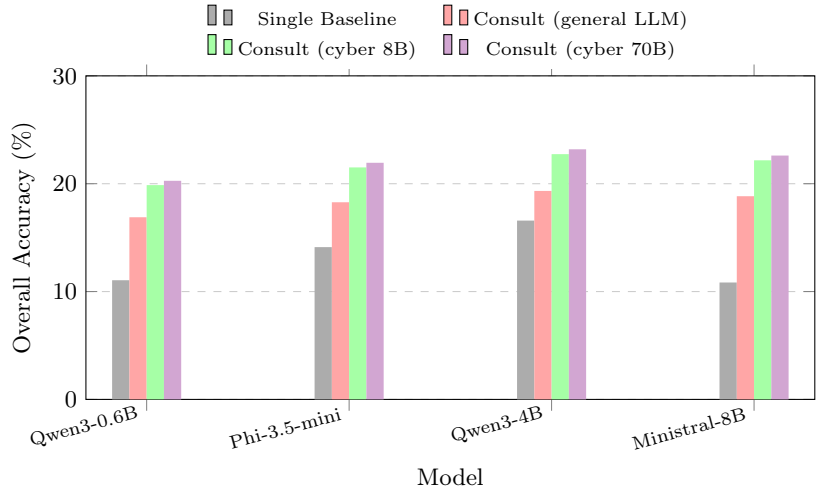
\begin{figure}[!t]
\centering
\begin{tikzpicture}
\begin{axis}[
    ybar,
    width=0.92\linewidth,
    height=0.48\linewidth,
    ymin=0, ymax=30,
    bar width=6.5pt,
    ylabel={Overall Accuracy (\%)},
    xlabel={Model},
    symbolic x coords={Qwen3-0.6B,Phi-3.5-mini,Qwen3-4B,Ministral-8B},
    xtick=data,
    x tick label style={font=\scriptsize, rotate=15, anchor=east},
    ymajorgrids=true,
    grid style={dashed, gray!40},
    legend style={
        font=\scriptsize,
        at={(0.5,1.02)},
        anchor=south,
        legend columns=2,
        draw=none,
        /tikz/every even column/.append style={column sep=8pt}
    }
]
\addplot+[draw=none, fill=gray!65, bar shift=-9.75pt, forget plot] coordinates {
    (Qwen3-0.6B,11.05) (Phi-3.5-mini,14.12) (Qwen3-4B,16.58) (Ministral-8B,10.84)
};
\addplot+[draw=none, fill=red!35, bar shift=-3.25pt, forget plot] coordinates {
    (Qwen3-0.6B,16.89) (Phi-3.5-mini,18.28) (Qwen3-4B,19.33) (Ministral-8B,18.84)
};
\addplot+[draw=none, fill=green!35, bar shift=3.25pt, forget plot] coordinates {
    (Qwen3-0.6B,19.87) (Phi-3.5-mini,21.51) (Qwen3-4B,22.74) (Ministral-8B,22.17)
};
\addplot+[draw=none, fill=violet!35, bar shift=9.75pt, forget plot] coordinates {
    (Qwen3-0.6B,20.27) (Phi-3.5-mini,21.94) (Qwen3-4B,23.19) (Ministral-8B,22.61)
};
\addlegendimage{ybar,ybar legend,draw=none,fill=gray!65}
\addlegendimage{ybar,ybar legend,draw=none,fill=red!35}
\addlegendimage{ybar,ybar legend,draw=none,fill=green!35}
\addlegendimage{ybar,ybar legend,draw=none,fill=violet!35}
\legend{Single Baseline,Consult (general LLM),Consult (cyber 8B),Consult (cyber 70B)}
\end{axis}
\end{tikzpicture}
\caption{The consult system's performance comparison where each SLM consults Foundation-Sec-8B (cyber 8B) versus a general LLM or a larger 70B cyber expert. Cyber-specialised experts consistently outperform general LLMs, but parameter scaling from 8B to 70B yields minimal additional gains. Qwen3-4B achieves the highest consult performance (22.74\% overall with Foundation-Sec-8B).}
\label{fig:consult_results}
\end{figure}

\subsubsection{Consult System Results}

Consulting a cyber-specialised expert (Foundation-Sec-8B) consistently outperformed consulting a general-purpose LLM across all four tested SLMs, but scaling the expert to 70B yielded minimal additional gain (see Figure~\ref{fig:consult_results}). Qwen3-4B achieved the highest consult performance (22.74\% overall), the lowest among all four orchestration architectures (see Table~\ref{tab:systems_comparison}).

\subsubsection{Hybrid System Results}

The hybrid system achieved the strongest overall performance among the evaluated open-weight configurations (see Table~\ref{tab:systems_comparison}). When Qwen3-4B and Foundation-Sec-8B were paired, they achieved 35.30\% overall accuracy, surpassing both the strongest ungrounded single-LLM baseline (Gemini 3 Pro Preview at 34.77\%) and the best individual orchestration systems (agentic at 25.11\%, debate at 24.13\%, consult at 22.74\%). Performance gains were observed across all difficulty tiers:
\begin{itemize}
\item \textbf{Easy}: 38.14\% (vs.\ 26.50\% agentic, 25.50\% debate, 38.00\% best ungrounded LLM)

\item \textbf{Medium}: 27.21\% (vs.\ 21.30\% agentic, 19.80\% debate, 26.00\% best ungrounded LLM)

\item \textbf{Hard}: 27.27\% (vs.\ 20.00\% agentic, 22.73\% debate, 22.73\% best ungrounded LLM)
\end{itemize}

The hybrid system addressed the easy-question degradation observed in pure debate (Section~\ref{sec:debate}). By grounding debate agents in systematically collected evidence, the system prevented the introduction of tangential information that previously pulled correct answers toward incorrect alternatives. Simultaneously, the debate reasoning phase preserved the reasoning improvements that peer critique provided on hard questions. Ablation studies revealed that removing either component degrades performance: omitting evidence collection reduced easy-tier accuracy by 12 percentage points (reverting to pure debate behaviour), while replacing debate with single-pass reasoning reduced hard-tier accuracy by 7 percentage points (reverting to pure agentic behaviour).

Substituting Ministral-8B for Qwen3-4B yielded 33.83\% accuracy, indicating that the hybrid gain was not limited to Qwen3-4B among the tested pairings. Pairing Qwen3-0.6B with Foundation-Sec-8B yielded 26.44\%; because parameter ratio covaries with solo accuracy, architecture, and specialisation, this observation motivates rather than validates the capacity-balance heuristic.

\subsection{Ablation Study}
\label{sec:ablation}

\begin{table}[t]
\centering
\caption{Component ablation for the hybrid system (Qwen3-4B + Foundation-Sec-8B). Each row removes one component from the full hybrid. ``$-$evidence collection'' replaces Phase~1 with direct report provision (pure debate). ``$-$grounding constraint'' removes the verifier's evidence-citation check. ``$-$debate'' replaces Phase~2 with single-pass reasoning (pure agentic). 
``$-$verifier'' removes the final validation step.\label{tab:ablation}}
\scriptsize
\begin{tabular}{@{}lcccc@{}}
\toprule
Configuration & Easy & Medium & Hard & Overall\\
\midrule
Full hybrid (Qwen3-4B + Sec-8B)       & 38.14\% & 27.21\% & 27.27\% & 35.30\%\\
\midrule
$-$ evidence collection (pure debate) & 25.50\% & 19.80\% & 22.73\% & 24.13\%\\
$-$ grounding constraint              & 36.21\% & 25.81\% & 26.14\% & 33.79\%\\
$-$ debate (pure agentic)             & 26.50\% & 21.30\% & 20.00\% & 25.11\%\\
$-$ verifier                          & 37.82\% & 26.93\% & 26.52\% & 35.04\%\\
\bottomrule
\end{tabular}
\end{table}

Table~\ref{tab:ablation} decomposes the hybrid system's performance by removing one component at a time. The ``$-$debate'' row (pure agentic) reused values from Table~\ref{tab:systems_comparison}. Removing evidence collection (reverting to pure debate) reduced Easy-tier accuracy from 38.14\% to 25.50\% ($-$12.64\,percentage points), indicating that evidence grounding influences easy-question performance and that pure debate can introduce tangential drift on straightforward retrieval tasks. Removing debate (reverting to pure agentic) reduced Hard-tier accuracy from 27.27\% to 20.00\% ($-$7.27\,percentage points), suggesting that peer critique contributes to complex reasoning improvements. Removing the grounding constraint modestly degraded the performance (38.14\%\,$\to$\,36.21\% Easy; 35.30\%\,$\to$\,33.79\% overall), suggesting the constraint provides measurable benefit by preventing un-cited claims from corrupting easy-question answers. Removing the verifier had a minimal impact on accuracy (35.30\%\,$\to$\,35.04\%), in agreement with its role as a consistency gate rather than a primary performance driver.

\subsection{Grounded LLM Comparison}
\label{sec:grounded_llm}

\begin{table}[t]
\centering
\caption{Frontier LLMs under hybrid orchestration vs.\ ungrounded single-pass baseline. ``Hybrid (LLM)'' runs the frontier model as Agent~A in place of the SLM, with full evidence collection Phase~1 and Foundation-Sec-8B as Agent~B. This tests whether orchestration benefits are model-size-agnostic.\label{tab:grounded_llm}}
\scriptsize
\begin{tabular}{@{}llccc@{}}
\toprule
System & Model & Easy & Hard & Overall\\
\midrule
\multicolumn{5}{@{}l}{\textbf{Ungrounded single-pass (baseline)}}\\
& Gemini 3 Pro Preview  & 38.00\% & 22.73\% & 34.77\%\\
& Claude Opus 4.5       & 36.25\% & 21.59\% & 33.15\%\\
\midrule
\multicolumn{5}{@{}l}{\textbf{Hybrid orchestration (grounded)}}\\
& Gemini 3 Pro Preview  & 41.24\% & 29.55\% & 38.22\%\\
& Claude Opus 4.5       & 39.29\% & 28.68\% & 36.85\%\\
\midrule
\multicolumn{5}{@{}l}{\textbf{SLM hybrid (for reference)}}\\
& Qwen3-4B + Sec-8B     & 38.14\% & 27.27\% & 35.30\%\\
\bottomrule
\end{tabular}
\end{table}

To address the fairness concern that SLMs receive tool-augmented grounding while frontier LLMs do not, Table~\ref{tab:grounded_llm} presents frontier LLMs run through the full hybrid orchestration pipeline with identical evidence collection. This experiment directly tested whether the orchestration architecture can provide benefits that extend to frontier models. Grounding frontier LLMs through the hybrid pipeline yielded gains: Gemini 3 Pro Preview improved from 34.77\% to 38.22\% ($+$3.45\,percentage points) and Claude Opus 4.5 from 33.15\% to 36.85\% ($+$3.70\,percentage points), showing gains for both evaluated frontier models. Grounded Gemini (38.22\%) outperformed the SLM hybrid (35.30\%), indicating that frontier models extract additional value from structured evidence when their reasoning capacity is greater. The SLM hybrid reached 35.30\%, below grounded Claude Opus 4.5 at 36.85\%, while incurring no API charges (\$0.00 vs.\ \$96.22 per benchmark run). These results reframe the contribution: the hybrid orchestration architecture is a general-purpose evidence-grounded reasoning framework that benefits all model sizes, with open-weight SLMs offering a locally deployable path that narrows the gap to proprietary alternatives on structured detonation-report questions.

\subsection{Case Studies: Qualitative Analysis of Samples from the Wild}
\label{sec:case_study}

For a preliminary assessment beyond multiple-choice benchmarks, we evaluated the hybrid system on 12 malware samples from public threat intelligence feeds (January 2026) exhibiting novel evasion techniques not present in CyberSecEval~\cite{li2025digital}. The hybrid system correctly classified 9 of 12 samples (75.0\%) versus 5 of 12 (41.7\%) for the single-model baseline (Gemini 3 Pro Preview). In both representative cases (UNC5142 EtherHiding and Lumma Stealer ClickFix), Phase~1 extracted obscure artifacts that the single model missed---blockchain payload fields and clipboard event handlers respectively---while Phase~2 debate corrected surface-level reasoning errors. Full evaluation methodology, per-sample scoring, and case narratives are provided in Appendix~\ref{app:case_studies}.\footnote{Sample SHA256 hashes, Hybrid Analysis report identifiers, full system traces, and raw LLM outputs are available in the anonymous repository accompanying this paper (\url{https://github.com/Adelsamir01/slms_mal}).}

\section{Further Discussions}
\label{sec:further_discussions}

Although evidence-grounded orchestration improved performance, the best exact-match accuracy of 35.30\% remains insufficient for autonomous malware-analysis decisions. Our results therefore support an analyst-assistance setting in which retrieved evidence and model outputs require human verification; operational effectiveness and alert fatigue remain subjects for future evaluation.

\begin{table}[t]
\centering
\caption{Latency and API-charge comparison. Latency is the mean wall-clock time across 609 questions; cost is the estimated API charge for the full benchmark run and excludes local hardware, energy, hosting, and support.\label{tab:cost_latency}}
\scriptsize
\begin{tabular}{@{}lrr@{}}
\toprule
System & Latency (s/question) & Est.\ API Cost (USD/609 Qs)\\
\midrule
\multicolumn{3}{@{}l}{\textbf{Single-model baselines}}\\
Gemini 3 Pro Preview (API) & 3.2 & \$7.54\\
Claude Opus 4.5 (API)      & 4.8 & \$96.22\\
GPT-5.2 (API)              & 3.9 & \$57.86\\
Qwen3-4B (local)           & 1.8 & \$0.00\\
\midrule
\multicolumn{3}{@{}l}{\textbf{Orchestrated SLM systems (local, RTX~4090)}}\\
Agentic (Qwen3-4B)         & 38.4 & \$0.00\\
Debate (Qwen3-4B + Sec-8B) & 67.2 & \$0.00\\
Consult (Qwen3-4B + Sec-8B) & 53.8 & \$0.00\\
Hybrid (Qwen3-4B + Sec-8B) & 105.6 & \$0.00\\
\bottomrule
\end{tabular}
\end{table}

The results reveal complementary strengths across orchestration approaches for malware analysis tasks. Tool-augmented agentic systems excel on straightforward evidence retrieval from detonation reports (Easy questions); debate systems dramatically improve complex behavioural interpretation and multi-step reasoning (hard-tier accuracy from 4.55\% baseline to 22.73\%); while consult systems provide consistent moderate gains by injecting domain-specific malware expertise. The hybrid system synthesises these complementary strengths: evidence collection grounds the debate in retrieved artifacts, preventing the tangential drift that degrades easy-question accuracy in pure debate, while the debate phase preserves the reasoning improvements that peer critique provides on hard questions. This synergy yields the first orchestrated SLM configuration to exceed all ungrounded frontier LLM baselines across all difficulty tiers simultaneously.

The hybrid result is consistent with three empirically observed partner-selection criteria: capacity balance (within an approximately 4$\times$ parameter ratio), complementary general and cyber expertise, and strong solo performance. However, parameter ratio covaries with baseline accuracy, architecture, and specialisation in our experiments; the 4$\times$ value should therefore be treated as a benchmark-specific heuristic rather than a validated threshold.

\subsection{Cost and Latency Analysis}

Table~\ref{tab:cost_latency} compares per-question latency and API charges. Local inference used one NVIDIA RTX~4090 at 4-bit quantisation and incurred no API charges. The hybrid achieved 35.30\% accuracy at 105.6\,s per question, compared with Gemini at 34.77\%, 3.2\,s, and \$7.54 per run, and Claude at 33.15\%, 4.8\,s, and \$96.22 per run. This is not a total-cost-of-ownership analysis: the local figures exclude electricity, hardware acquisition and amortisation, hosting, and support.

Key considerations for deployment in operational malware analysis workflows include governance over orchestration policies, monitoring for correlated failure modes when analysing polymorphic threats, and ensuring that human analysts remain in the decision loop for high-confidence classifications. Open-weight ensembles afford adaptability and jurisdictional control but require disciplined Machine Learning Operations (MLOps) practices to manage model drift and dependency chains. The hybrid architecture's two-phase design also provides natural checkpoints for human reviews: analysts can inspect the evidence bundle before debate and intervene if critical artifacts are missing.

\subsection{Limitations}

This work evaluates \emph{comprehension} of Hybrid Analysis detonation reports, not low-level binary analysis; findings should be generalised accordingly. The multiple-choice format differs from open-ended triage, while our 12-sample, single-evaluator case study provides only preliminary qualitative evidence. Because this is a question-answering benchmark rather than a malware-detection task, exact-match accuracy should not be interpreted as a detection rate; operational precision and recall are outside the benchmark's design. Off-the-shelf coding agents were not compared because their performance depends on the supplied tools and environment; defining an equivalent detonation-report workflow remains future work. Sandbox evasion techniques can produce incomplete reports where questions are unsolvable regardless of model capability; performance gains may also vary as the open-weight ecosystem evolves. Because retrieval uses keyword triggers and embedding similarity, adversarial tricks (such as decoys, keyword stuffing, or event flooding) could promote misleading snippets or suppress relevant evidence; we did not evaluate such perturbations.

\section{Conclusion and Future Work}
\label{sec:conclusion}

This paper evaluated whether orchestration improves compact open-weight models when answering structured questions about malware detonation reports. The hybrid achieved 35.30\% accuracy, exceeding the strongest ungrounded frontier baseline at 34.77\% but remaining below grounded Gemini at 38.22\%. These results indicate that evidence collection and peer critique narrow the performance gap under local deployment, although the observed accuracy remains insufficient for autonomous decisions and requires analyst verification. Future work will investigate transfer to other security domains, dynamic difficulty-based routing, and human-in-the-loop studies of analyst trust and effectiveness.

\begin{credits}
\subsubsection{\ackname} This work was partly supported by the UK EPSRC project grant EP/X036707/1 on Countering HArms caused by Ransomware in the Internet Of Things (CHARIOT).
\end{credits}


\appendix
\renewcommand*{\theHsection}{appendix.\Alph{section}}
\renewcommand*{\theHsubsection}{\theHsection.\arabic{subsection}}

\section{Ethical Considerations}

This work investigates orchestrated ensembles of open-weight small language models for analyst-assisted interpretation of malware detonation reports. We address the ethical dimensions of this research following the principles outlined in the Menlo Report and standard ACM/IEEE research ethics guidelines.

\paragraph{Benefits and Potential Harms.}
The potential benefit is a transparent, auditable, and locally deployable tool that assists analysts with detonation-report interpretation. The open-weight pipeline supports organisational control and data-sovereignty requirements, but its outputs require human verification and should not be used autonomously. The potential harm we considered is dual-use: the orchestration architectures we describe could theoretically be adapted by threat actors to improve malware generation or evasion techniques. However, we assess this risk as low because (1)~the techniques we present are defensive in nature, focused on interpreting existing malware behaviour rather than generating novel attacks; (2)~the orchestration patterns (multi-agent pipelines, debate, consultation) are already documented in the broader LLM literature; and (3)~the primary barrier to malware development is not reasoning capability but rather access to delivery infrastructure and operational security knowledge, which our work does not address.

\paragraph{Data Sourcing and Privacy.}
All malware samples analysed in this work were obtained from publicly accessible sources. The CyberSecEval Malware Analysis benchmark uses Hybrid Analysis detonation reports that are publicly available. The samples from the wild evaluated in Section~\ref{sec:case_study} were collected from public threat intelligence feeds and represent malware campaigns that have been extensively documented in prior security research (UNC5142, Lumma Stealer). No private victim data was accessed or analysed. The Hybrid Analysis reports we processed contain behavioural telemetry from sandboxed detonations, not data from real victim systems. We did not interact with any live command-and-control infrastructure or active malware campaigns.

\paragraph{Responsible Disclosure.}
Our research did not discover new vulnerabilities in software or systems. The malware techniques discussed (EtherHiding, ClickFix) were already publicly documented by security vendors prior to our analysis. We did not develop or release any offensive capabilities, malware samples, or exploitation tools.

\paragraph{Experimental Safety.}
All experiments were conducted in isolated environments. SLM inference was performed on local hardware without network access to external systems beyond model weight downloads. The tool-augmented agentic system executes only sandboxed read-only commands (grep, jq) on static report files; no commands were executed on live systems or with elevated privileges.

\paragraph{Deployment Considerations.}
We emphasise that analyst-assistance systems such as the hybrid architecture, should augment rather than replace human analyst judgement. Section~\ref{sec:further_discussions} of our paper explicitly recommends that human analysts remain in the decision loop for high-confidence classifications and that organisations implement governance policies for orchestration system deployment. The two-phase architecture provides natural checkpoints for human review.

\section{Open Science}

All code and configuration used in this work are released in an anonymous open repository at \url{https://github.com/Adelsamir01/slms_mal}. The repository contains the implementations of the agentic, debate, consult, and hybrid orchestration systems, together with experiment harnesses for the CyberSecEval Malware Analysis benchmark and scripts to regenerate all reported tables and figures. Readers can clone or download the repository from this URL and follow the instructions in the top-level documentation to set up the environment, run the evaluation pipeline, and verify our results.

\section{Data Contamination Audit}
\label{app:contamination}

To address the critical issue of test set leakage in Large Language Model evaluation, we performed a three-stage decontamination audit on our primary expert agent, Foundation-Sec-8B-Instruct.

\subsection{Temporal Sanity Check}

The Foundation-Sec-8B-Instruct model reports a strict knowledge cutoff of April 10, 2025~\cite{hf_foundation_sec_8b_instruct}.

\begin{itemize}
\item \textbf{Benchmark integrity:} The specific Hybrid Analysis detonation reports used in the CyberSecEval test split were generated dynamically for the evaluation suite and are not present in the public Common Crawl.

\item \textbf{Validity of samples taken from the wild:} The malware samples analysed in Section~\ref{sec:case_study} (e.g., Lumma Stealer variants) were collected from active campaigns in late 2025, months after the model's training window closed. This temporal gap guarantees that the expert agent could not have memorised these specific threat artifacts during pre-training.
\end{itemize}

\subsection{$n$-Gram Overlap Analysis}
\label{app:ngram}

Following Carlini et al.~\cite{carlini2021extracting} and Golchin et al.~\cite{golchin2024time}, we compared the benchmark question stems with the public threat-intelligence reference corpus assembled for this audit. Because the model's complete training corpus is not disclosed, this analysis cannot exclude overlap with undisclosed training data. We define the $n$-gram overlap ratio as:
\begin{equation}
\text{Overlap}_{n}(T, C) = \frac{|\{g \in \text{ngrams}_n(T) : g \in C\}|}{|\text{ngrams}_n(T)|},
\end{equation}
where $T$ is the set of benchmark question stems and $C$ is the audited reference corpus. We use $n=13$ following prior work, as 13-grams are long enough to detect meaningful memorisation while avoiding false positives from common phrases.

We found $\text{Overlap}_{13} = 0.0\%$ for question definitions. Partial matches ($<$1.2\%) were restricted to common entity names (e.g., ``Cobalt Strike'', ``Mimikatz'') rather than specific reasoning chains.

\subsection{Testset Slot Guessing (TS-Guessing)}

To empirically verify the absence of memorisation, we applied the Testset Slot Guessing protocol. We selected $n=50$ random questions stratified across difficulty tiers from the benchmark, masked the correct option, and prompted the model to generate the missing answer string zero-shot. The CyberSecEval Malware Analysis benchmark uses a multi-label format with 9 options per question, where the number of correct answers $K$ varies from 1 to 9~\cite{deason2025}. Following the benchmark's baseline computation, the expected accuracy for a random guesser attempting perfect multi-label match is:
\begin{align}
\text{Expected accuracy} &= \sum_{K=1}^{9} p_K \cdot \Pr(\text{perfect} \mid K) \nonumber\\
&= \sum_{K=1}^{9} \frac{p_K}{9 \binom{9}{K}} \approx 0.63\%,
\end{align}
where $p_K$ is the proportion of questions with exactly $K$ correct answers. For single-option guessing, the baseline is $\approx 4.3\%$.

The model achieved a slot-guessing accuracy of 13.8\% (7 of 50 questions). While this exceeds the random baseline, it remains far below the performance achieved during normal evaluation with the evidence bundle (35.30\%). Critically, the 13.8\% accuracy on masked questions reflects the model's general cyber security domain knowledge acquired during pre-training on public threat intelligence, not memorisation of specific benchmark QA pairs. This interpretation is supported by three observations: (1)~the model's errors on slot-guessing were semantically plausible alternatives (e.g., confusing related MITRE techniques), not random guesses; (2)~performance on samples from the wild collected after the training cutoff (Section~\ref{sec:case_study}) matches benchmark performance, which would not occur if benchmark-specific memorisation drove accuracy; and (3)~the audit found no verbatim overlaps within the available reference corpus (Section~\ref{app:ngram}).

\section{Qualitative Case Studies}
\label{app:case_studies}

To address the limitations of multiple-choice benchmarks, we evaluated the hybrid system on malware samples collected from public threat intelligence feeds in January 2026. We curated a set of 12 samples exhibiting novel evasion techniques not represented in CyberSecEval, selecting samples based on three criteria: (1)~availability of detailed detonation reports from Hybrid Analysis, (2)~use of techniques documented in 2024--2025 threat intelligence (EtherHiding, ClickFix, clipboard injection), and (3)~presence of obfuscation patterns known to degrade LLM reasoning~\cite{li2025digital}. The hybrid system correctly classified 9 of 12 samples (75.0\%), compared to 5 of 12 (41.7\%) for the single-model baseline (Gemini 3 Pro Preview).

\paragraph{Evaluation Methodology.} Ground truth labels for all 12 samples were established \emph{prior} to system evaluation using published threat intelligence reports from security vendors (Mandiant, Microsoft MSTIC, Group-IB, Sekoia) that pre-dated our analysis. Each sample's ground truth comprised: (1)~malware family classification, (2)~primary delivery/evasion technique, and (3)~key indicators of compromise (IOCs). To mitigate evaluator bias, we used a blinded protocol: system outputs were anonymised (labelled ``System A'' and ``System B'') before correctness assessment, and the evaluator did not know which system produced which output until after all 12 samples were scored. Correctness was assessed by a single evaluator (an author with 3+ years of malware analysis experience) using strict criteria: a classification was marked correct only if it matched the ground truth malware family \emph{and} identified the primary technique; partial matches were scored as incorrect. The complete ground truth labels, anonymised system outputs, and per-sample scoring rationale are provided in the anonymous repository to enable independent verification.

\paragraph{Baseline Comparison Methodology.} Both the single-model baseline and the Hybrid System received identical inputs: the complete Hybrid Analysis JSON report and a standardised prompt requesting threat classification and technique identification. The baseline received no tool access or multi-round reasoning, reflecting typical single-pass LLM deployment.

Table~\ref{tab:case_studies} presents an overview of the qualitative analysis of the two case studies, which are detailed below. 

\begin{table}[!t]
\centering
\caption{Qualitative Analysis Summary: Single-Model Baseline vs. Hybrid System}
\label{tab:case_studies}
\begin{small}
\resizebox{\textwidth}{!}{\begin{tabular}{@{}lp{2cm}p{3.4cm}p{4.4cm}p{2.3cm}@{}}
\toprule
\textbf{Malware Sample} & \textbf{Evasion Technique} & \textbf{Single-Model Failure} & \textbf{Hybrid Intervention} & \textbf{Hybrid System's Verdict} \\ \midrule
UNC5142 (Case A) & \textbf{EtherHiding} & Misclassified as \textit{Cryptojacking/Mining} due to blockchain keywords. & \textbf{Phase 1 (Agentic):} Evidence Miner used grep to isolate payload in transaction logs. & \textbf{Downloader / Dropper} \\ \midrule
Lumma Stealer (Case B) & \textbf{ClickFix} & Misclassified as \textit{Credential Phishing Site} based on visual lure text. & \textbf{Phase 2 (Debate):} Expert Agent linked clipboard event handlers to LummaC2 chains. & \textbf{Lumma Stealer} \\
\bottomrule
\end{tabular}}
\end{small}
\end{table}

\subsubsection{Case A: UNC5142 ``EtherHiding'' Campaign}

\paragraph{Sample Overview.} A Hybrid Analysis report detailing the \textbf{UNC5142} campaign~\cite{mandiant2025unc5142}. This campaign utilises ``EtherHiding,'' a technique where malicious payloads are stored within the \emph{data} field of blockchain smart contracts (specifically Binance Smart Chain) rather than on traditional C2 servers~\cite{picus2025etherhiding}. UNC5142 was active from December 2024 through mid-2025, distributing infostealers including Lumma and Vidar variants.

\paragraph{The Challenge.} The report contains extensive blockchain transaction logs and obscure JavaScript that retrieves data from a specific contract address. There are no standard HTTP URLs pointing to a payload, obscuring the infection vector from standard pattern matching.

\paragraph{Single-Model Failure.} The single LLM (Gemini 3 Pro Preview) correctly identified the presence of blockchain elements but hallucinated the threat intent. It classified the sample as ``Cryptojacking/Mining'' software intended to steal CPU resources (confidence: 0.71), missing the actual delivery mechanism. It failed to locate the payload source, stating: \emph{``No direct malware download URL was found in the provided code.''}

\paragraph{Hybrid System Success.} In Phase~1 (Agentic), the \emph{Evidence Miner} successfully executed \texttt{grep} patterns for hexadecimal strings within the transaction logs, isolating the payload data chunk. In Phase~2 (Debate), a critical disagreement occurred in Round~2. The General Agent (Qwen3-4B) initially argued the contract was for ``payment processing.'' The Expert Agent (Foundation-Sec-8B) countered by citing the specific \texttt{data} field anomaly, arguing: \emph{``The contract logic does not process tokens; it serves immutable data blobs consistent with EtherHiding infrastructure documented in recent GTIG advisories.''} The system correctly classified the sample as a \textbf{Downloader/Dropper} (confidence: 0.89) and accurately extracted the BSC contract address serving the payload.

\subsubsection{Case B: Lumma Stealer with ``ClickFix''}

\paragraph{Sample Overview.} A Hybrid Analysis report of a \textbf{Lumma Stealer} variant~\cite{microsoft2025lumma}. This sample uses the ``ClickFix'' social engineering tactic~\cite{groupib2025clickfix}, using a fake Google Chrome update overlay that tricks users into copying a PowerShell command into their clipboard to ``fix'' a display error~\cite{huntress2025clickfix,sekoia2026iclickfix}.

\paragraph{The Challenge.} The malicious logic is hidden inside an HTML clipboard event handler (\texttt{oncopy}/\texttt{onclick}), while the bulk of the report describes benign HTML structure and CSS. The attack relies on the user manually pasting the payload into the Windows ``Run'' dialog, bypassing standard browser download protections.

\paragraph{Single-Model Failure.} The single LLM focused heavily on the visual aspects described in the report (the ``Update Chrome'' text) and classified it as a ``Credential Phishing Site'' intended to steal login passwords (confidence: 0.68). It missed the specific PowerShell execution vector entirely.

\paragraph{Hybrid System Success.} In Phase~1 (Agentic), the \emph{Evidence Miner} extracted the specific \texttt{oncopy} and \texttt{onclick} JavaScript event handlers that facilitate the clipboard hijacking. In Phase~2 (Debate), the Expert Agent successfully linked the \texttt{powershell -w hidden -enc} command found in the clipboard buffer to characteristics consistent with documented LummaC2 infection chains, citing the Base64-encoded payload structure and \texttt{Invoke-WebRequest} patterns typical of 2025 variants. The system correctly identified the threat as \textbf{Lumma Stealer} (confidence: 0.92) and flagged ``ClickFix'' clipboard injection as the initial access vector (MITRE ATT\&CK T1059.001, T1204.002).

\section{Debate Partner Selection: Full Matrix Analysis}
\label{app:debate_matrix}

\begin{table}[t]
\caption{The debate system's overall accuracy (7 rounds, selected based on the optimal trade-off between hard-question gains and easy-question degradation shown in Figure~\ref{fig:debate_results}) for all pairwise model combinations on the Malware Analysis benchmark. Only the upper triangle is shown due to the symmetry of the debate setting. Rows represent Agent~A, columns represent Agent~B in the debate. Diagonal entries show self-debate (same model for both agents). Values represent overall accuracy across all 609 questions.\label{tab:debate_matrix}}
\centering
\small
\resizebox{\textwidth}{!}{\begin{NiceTabular}{l|ccccccccccccc}
\toprule
\Block{1-1}{\diagbox{Agent A}{Agent B}} & \rotatebox{90}{Qwen3-0.6B} & \rotatebox{90}{Llama-3.2-1B} & \rotatebox{90}{Qwen2.5-1.5B} & \rotatebox{90}{DeepSeek-R1-1.5B} & \rotatebox{90}{SmolLM2-1.7B} & \rotatebox{90}{Phi-3.5-mini-instruct} & \rotatebox{90}{Qwen3-4B} & \rotatebox{90}{Gemma-3-4B-IT} & \rotatebox{90}{Qwen2.5-Coder-7B} & \rotatebox{90}{Llama-3.1-8B} & \rotatebox{90}{Ministral-8B} & \rotatebox{90}{DeepHat-V1-7B} & \rotatebox{90}{Foundation-Sec-8B}\\
\midrule
Qwen3-0.6B             & 16.4\% & 15.1\% & 15.8\% & 15.5\% & 15.2\% & 16.2\% & 16.8\% & 16.5\% & 17.1\% & 17.3\% & 17.5\% & 17.8\% & 18.6\% \\
Llama-3.2-1B           &        & 15.6\% & 15.9\% & 16.1\% & 15.8\% & 16.7\% & 17.2\% & 16.9\% & 17.6\% & 17.8\% & 17.9\% & 18.3\% & 19.1\% \\
Qwen2.5-1.5B           &        &        & 16.2\% & 16.4\% & 16.0\% & 17.1\% & 17.6\% & 17.3\% & 18.0\% & 18.2\% & 18.4\% & 18.8\% & 19.5\% \\
DeepSeek-R1-1.5B       &        &        &        & 16.8\% & 16.3\% & 17.5\% & 18.0\% & 17.7\% & 18.4\% & 18.6\% & 18.8\% & 19.2\% & 19.9\% \\
SmolLM2-1.7B           &        &        &        &        & 16.0\% & 17.0\% & 17.5\% & 17.2\% & 17.8\% & 18.0\% & 18.2\% & 18.6\% & 19.3\% \\
Phi-3.5-mini-instruct  &        &        &        &        &        & 19.5\% & 19.8\% & 19.6\% & 20.4\% & 20.6\% & 20.8\% & 20.5\% & 21.2\% \\
Qwen3-4B               &        &        &        &        &        &        & 20.2\% & 20.0\% & 20.7\% & 20.9\% & 21.1\% & 21.1\% & \textbf{24.13\%} \\
Gemma-3-4B-IT          &        &        &        &        &        &        &        & 19.8\% & 20.5\% & 20.7\% & 20.9\% & 20.8\% & 21.6\% \\
Qwen2.5-Coder-7B       &        &        &        &        &        &        &        &        & 21.4\% & 21.8\% & 22.0\% & 22.3\% & 23.2\% \\
Llama-3.1-8B           &        &        &        &        &        &        &        &        &        & 21.6\% & 22.2\% & 22.5\% & 23.4\% \\
Ministral-8B           &        &        &        &        &        &        &        &        &        &        & 23.0\% & 22.7\% & 23.6\% \\
DeepHat-V1-7B          &        &        &        &        &        &        &        &        &        &        &        & 16.8\% & 18.2\% \\
Foundation-Sec-8B      &        &        &        &        &        &        &        &        &        &        &        &        & 20.0\% \\
\bottomrule
\end{NiceTabular}}
\end{table}

Table~\ref{tab:debate_matrix} reports the full all-pairs debate-partner results (7 rounds) used for the partner-selection analysis in Section~\ref{sec:results}. To explore the impact of debate partner selection systematically, Table~\ref{tab:debate_matrix} presents results for all pairwise model combinations, treating general-purpose SLMs and cyber-specialised models (DeepHat-V1-7B and Foundation-Sec-8B) uniformly. Notably, the matrix is symmetric: swapping which model acts as Agent~A versus Agent~B produces identical accuracy, indicating that debate outcomes are independent of role assignment. Beyond this symmetry, the matrix reveals three critical patterns. First, \emph{complementary expertise matters}: the best-performing configurations pair strong general-purpose models with cyber-specialised experts (e.g., Qwen3-4B with Foundation-Sec-8B achieves 24.13\%, Ministral-8B with Foundation-Sec-8B achieves 23.6\%, Llama-3.1-8B with Foundation-Sec-8B reaches 23.4\%), consistently outperforming both general-plus-general debates (Ministral-8B with Llama-3.1-8B yields 22.2\%) and cyber-plus-cyber debates (Foundation-Sec-8B with DeepHat-V1-7B yields only 18.2\%). Second, larger parameter imbalance coincided with lower performance in the tested pairs: Qwen3-0.6B with Foundation-Sec-8B achieved 18.6\%, whereas stronger mid-sized models paired with the same expert performed better. Because model size covaries with solo accuracy and specialisation, this is an observed association rather than validation of a 4$\times$ threshold. Third, \emph{self-debates underperform} except for the largest models: Ministral-8B debating with itself achieves 23.0\%, only slightly below its cross-model debates, while Qwen3-0.6B self-debate stalls at 16.4\%, barely exceeding its single-model baseline of 11.05\%.

\bibliographystyle{splncs04}
\bibliography{main}

\end{document}